\documentclass{article}
\usepackage[english]{babel}
\usepackage{amssymb}
\usepackage{amsfonts}
\usepackage{amsmath}
\usepackage{graphicx}
\usepackage{epsfig}
\usepackage{cite}
  \newcommand{\cB}{{\cal B}}

\newcommand{\cM}{{\cal M}}

\newcommand{\bP}{{\mathbf P}}  
\newcommand{\bR}{{\mathbf R}}

  \newcommand{\bZ}{{\mathbf Z}}

\newcommand{\bx}{{\mathbf x}}

\newcommand{\be}{\begin{equation}} \newcommand{\ee}{\end{equation}}
\newcommand{\bea}{\begin{eqnarray}} \newcommand{\eea}{\end{eqnarray}}
\newcommand{\beann}{\begin{eqnarray*}}  \newcommand{\eeann}{\end{eqnarray*}}
\newcommand{\bfig}{\begin{figure}} \newcommand{\efig}{\end{figure}}
\newcommand{\ba}{\begin{array}} \newcommand{\ea}{\end{array}}
\newcommand{\bcen}{\begin{center}} \newcommand{\ecen}{\end{center}}
\newcommand{\btab}{\begin{tabular}} \newcommand{\etab}{\end{tabular}}

\def\tr{\operatorname{tr\:}}

\def\None{{\cal N}= 1 } \def\Ntwo{{\cal N}=2 } \def\Nfour{{\cal N}=4 }

  \def\Qt{\widetilde{Q}}

\def\qt{\widetilde{q}}

\newtheorem{Proposition}{Proposition}[section]

\newtheorem{Theorem}{Theorem}[section]
\newtheorem{Lemma}{Lemma}[section]
\newtheorem{Corrolary}{Corrolary}[section]

\newcommand{\bp}{\begin{Proposition}}   \newcommand{\ep}{\end{Proposition}}
\newcommand{\bt}{\begin{Theorem}}   \newcommand{\et}{\end{Theorem}}
\newcommand{\bl}{\begin{Lemma}}     \newcommand{\el}{\end{Lemma}}
\newcommand{\bc}{\begin{Corrolary}} \newcommand{\ec}{\end{Corrolary}}

\def\psibar{\overline \psi}
\def\qbar{\overline q}

\def\qbart{\overline{\widetilde q}}

\title{\bf \huge Alternative large $N_c$ baryons and holography}
\author{ Carlos Hoyos, Andreas Karch \vspace{0.4cm} \\
\small Department of Physics,University of Washington\\
\small ~\,Seattle, WA 98915-1560, USA\\
\small  ~\,E-mail: \tt{choyos@phys.washington.edu, karch@phys.washington.edu}}

\begin{document}
\maketitle

\centerline{\bf Abstract:}
In gauge theories in the limit of a large number $N_c$
of colors  baryons
are usually described as heavy solitonic objects with mass of order $N_c$.
We discuss an alternative large $N_c$
description both directly in the field theory as well as
using holography.
In this alternative large $N_c$ limit at
least some of the baryons behave like mesons, that is
they stay light even at large $N_c$ and their interactions vanish
in that limit. For $N_c=3$ these alternative
large $N_c$ baryons are equivalent to the standard baryons. In
the holographic description it is manifest that the Regge slopes
of mesons and alternative baryons are degenerate.

\vskip2cm

\section{Introduction}

The study of baryons in QCD in the limit of large number $N_c$ of colors
has a long history starting with \cite{Witten:1979kh}. While mesons
are bound states of a single quark-antiquark pair for any value of $N_c$,
baryons are typically thought of as bound states of $N_c$ quarks and as
such are heavy in the large $N_c$ limit. While a meson/meson interaction
goes to zero at large $N_c$, the baryon meson interaction remains of order 1
and as such baryons are somewhat complicated to study even at large $N_c$.
In this paper we discuss an alternative large $N_c$ limit first proposed in \cite{Corrigan:1979xf}
and rediscovered in \cite{Armoni:2003gp, Armoni:2004ub} in which
at least some of the baryons behave like mesons in the large $N_c$ limit.
For $N_c=3$ the model reduces to QCD as the standard large
$N_c$ limit does.

The basic fact underlying the construction of these alternative
baryons is that for $N_c=3$ the antisymmetric and
the fundamental representations
are conjugate, so a gauge theory with quarks in the 2-index
antisymmetric tensor representation can also be thought as a
large-$N_c$ limit of QCD, different from the 't Hooft or the
Veneziano limits, but sharing with the last that fermion loops
are not suppressed.

The large-$N_c$ limit of gauge theories has a number of
interesting features that have been useful to improve our understanding of
the strongly coupled regime. One of the most interesting is the equivalence
between common sectors of theories with a different field content, where the
identification of the common sector is made with the help of global
symmetries (e.g.\cite{Kovtun:2003hr}). In particular, it is possible to show a relation between
supersymmetric and non-supersymmetric theories, so results from
non-perturbative techniques that usually apply only to supersymmetric
theories can be translated to their partners. An example of this are
orientifold theories, where the gaugino is mapped into fermions in the
antisymmetric representation \cite{Armoni:2003gp, Armoni:2004ub,Armoni:2003fb} and so gives a realization of the alternative baryons. This orientifold
theory reduces to QCD at $N_c=3$, but at the same time is equivalent
to SUSY QCD at large $N_c$.
Some non-perturbative results from the
supersymmetric theory can then be extrapolated to
QCD \cite{Armoni:2005wt, Armoni:2004uu}.

A recent result of the orientifold equivalence is a degeneracy between
the Regge slopes of mesons and baryons \cite{Armoni:2009zq}.
A degeneracy of this kind has been observed experimentally \cite{Selem:2006nd, Afonin:2007bm},
a possible explanation is the existence of diquark states that
form a bound state with a third quark (e.g. \cite{Selem:2006nd}).
This has the problem of the non-observation of tetraquarks,
although exotic states are quite difficult to identify in general.
The degeneracy is specially difficult to explain from the
point of view of the usual large-$N_c$ limits, where baryons
are always formed with $N_c$ quarks so they are very heavy,
non-perturbative objects. However, in the orientifold theory there is an additional set of operators that resemble the diquark states and
that map to baryons when $N_c=3$. They are formed with the antisymmetric
and two fundamental fermions
$$
 \psibar^{\; [ij]}q_i q_j\,.
$$
These are the objets that have the same Regge slope as the mesons.
While this gives a nice explanation why the Regge slope of mesons
and alternative baryons can be expected to be degenerate, one has
to keep in mind that in the orientifold theory not all of the
standard $N_c=3$ baryons turn into alternative baryons, but
some have to be extrapolated to conventional large $N_c$ baryons. An early analysis of the QCD phenomenology of this model and its differences with the usual large $N_c$ limit was made in \cite{Kiritsis:1989ge}.

Of course, the model presented above is not the only possibility, there are many different orientifold large $N_c$ limits of QCD; for every flavor one has the choice of extrapolating it either as an anti-symmetric two index tensor or a fundamental \footnote{Up to five flavors, in order to preserve asymptotic freedom.}. It is even possible to construct a chiral version by promoting the Weyl components of the Dirac spinor to different representations \cite{Ryttov:2005na,Sannino:2007yp}. So in principle for every baryon that has at least two different flavors of quarks, one in the fundamental and the other in the antisymmetric, one can find a large $N_c$ limit in which it becomes an alternative baryon with a string tension equal to that of the corresponding meson.

Our purpose in this paper is to study the spectrum
of alternative baryons and their relation both to the light
mesons and the conventional baryons at strong coupling
using a holographic setup \cite{firstadscft}.
Holographic constructions give a geometrical, weakly coupled
description of strongly coupled gauge theories in the large-$N_c$ limit, so
they are a natural ground to test large-$N_c$ equivalences at strong coupling.
As we are interested mostly in the conceptual framework
and the equivalence between mesons and alternative baryons,
we can choose to study a very symmetric situation where
the understanding of the holographic dictionary is well
established and non-perturbative effects are under control at the
cost of studying a theory quite distinct from QCD.
The simplest choice is a \Ntwo theory with antisymmetric and fundamental matter that is large-$N_c$ equivalent to an \Nfour superconformal theory. The holographic dual corresponds to type IIB string theory on $AdS_5\times \bR\bP^5$ geometry with probe D7 branes and an orientifold O7 plane wrapping an $\bR\bP^3\subset \bR\bP^5$ cycle. In addition to mesons and light baryons it is also interesting to study other objects, including heavy baryons and heavy `mesons', that are Pfaffian operators with $\sim N_c/2$ fields. For $N_c$ odd they look as
$$
\epsilon^{i_1 i_2\dots i_{N_c-2} i_{N_c-1} j} \psi_{i_1 i_2} \cdots \psi_{i_{N_c-2} i_{N_c-1}} q_j\,.
$$
When $N_c=3$ this operator is a bilinear and it corresponds to a meson operator.
We will see that there is indeed a degeneracy in the high energy spectrum of mesons and light baryons and find an interesting hierarchy between mesons, light baryons, heavy mesons and heavy baryons at strong coupling.

\section{Field theory preliminaries}

The orientifold field theory consists of an $SU(N_c)$ gauge theory with a Dirac fermion $\psi_{ij}$ in the two-index antisymmetric representation of the group. It is possible also to introduce a small number $N_f\ll N_c$ of fermions in the fundamental representation $q_i$. In the large-$N_c$ limit, this theory is equivalent to a supersymmetric theory in the common sector\footnote{We are assuming the theories to be defined in flat non-compact spacetime, in other situations spontaneous breaking of global symmetries could spoil the equivalence \cite{Unsal:2006pj}.} \cite{Armoni:2003gp,Armoni:2004ub}. When $N_c=3$, the orientifold theory is actually QCD with $N_f+1$ flavors, since the antisymmetric and the fundamental representations of $SU(3)$ are conjugate
\begin{equation}
\psibar^{\; ij}={1\over 2}\epsilon^{ijk}\qt_k\,.
\end{equation}
This has interesting consequences for the large-$N_c$ limit of the theory, since the usual hierarchy of operators is modified with respect to the usual 't Hooft or Veneziano limits \cite{'tHooft:1973jz, Veneziano:1976wm}. In these limits the quarks are always in the fundamental representation, so in the flavor sector one can distinguish `light' operators with a small number of quarks like mesons from `heavy' operators with a large number of quarks, like baryons. The operators associated to mesons and baryons are
\begin{equation}\label{oper}
M_1= \qbar^{\;i} q_i, \ \ B_1 = \epsilon^{i_1 i_2\cdots i_{N_c}} q_{i_1} q_{i_2}\cdots q_{i_{N_c}}\,.
\end{equation}

In the orientifold limit, in addition to \eqref{oper}, there are $N_c=3$ mesons and baryons involving the antisymmetric flavor $\qt$ that map differently. It is possible to have light `baryons' made of meson-like objects involving an antisymmetric fermion
\begin{equation}\label{baryon}
B_2 = \psibar^{\; ij}q_i q_j\,.
\end{equation}
In addition to light baryons, there are also heavy `mesons' made of Pfaffian-like objects. For $N_c$ even they are made entirely out
of the antisymmetric fermions, while for $N_c$
odd they can be constructed including a fundamental fermion
\begin{equation}\label{meson}
M_2 =
\left
\{ \begin{array}{ll}
\epsilon^{i_1 i_2\dots i_{N_c-1} i_{N_c} }
\psi_{i_1 i_2} \cdots \psi_{i_{N_c-1} i_{N_c}} \,& \mbox{ for } N_c
\mbox{ even } \\
\epsilon^{i_1 i_2\dots i_{N_c-2} i_{N_c-1} j} \psi_{i_1 i_2} \cdots
\psi_{i_{N_c-2} i_{N_c-1}} q_j\, &\mbox{ for } N_c
\mbox{ odd. } \\
\end{array} \right .
\end{equation}
When $N_c=3$, the operator \eqref{meson} becomes
\begin{equation}
M_2= \epsilon^{ijk}\psi_{ij}q_k = \qbart^{\; k} q_k\,.
\end{equation}
Notice that the additional mesons at $N_c=3$ are related to the enhancement of the flavor group
$$
SU(N_f)_V\times SU(N_f)_A\times U(1)_V\times U(1)_A\times U(1) \to U(1)_B\times SU(N_f+1)_V\times SU(N_f+1)_A\,.
$$
The $U(1)_A\times U(1)$ subgroup of the large-$N_c$ theory corresponds to the vector $U(1)$ that rotates the antisymmetric fermion and to an anomaly-free combination of the axial $U(1)$ symmetries that rotates fundamental and antisymmetric flavors \cite{Armoni:2005wt}. When $N_c=3$, a combination of $U(1)$
and $U(1)_V$ gives the baryon number while the rest of
components enter into the non-Abelian flavor group. We can generalize
this standard $N_c=3$ definition of baryon number as a linear combination
of $U(1)$ and $U(1)_V$ to any $N_c$. If we assign baryon number
$\frac{1}{N_c}$ to the $q_i$ and $\bar{\psi}^{ij}$, and then correspondingly
baryon number $-\frac{1}{N_c}$ to $\bar{q}^i$ and $\psi_{ij}$ the mesons
$M_1$ indeed have baryon number 0, while both conventional and alternative
baryons $B_1$ and $B_2$ have baryon number 1. The Pfaffian-like
objects $M_2$ will have non-zero baryon number in general, but turn
into mesons with baryon number 0 at $N_c=3$.

The relation between \eqref{baryon} and meson operators has been made more manifest by using the equivalence with a supersymmetric theory \cite{Armoni:2009zq}. An external quark anti-quark pair is introduced to describe a meson. There is a string of flux between the pair, so a supersymmetric transformation of this object will introduce a gaugino operator $\lambda$ along the string, roughly
$$
\qbar e^{i\int A} q \to \qbar \lambda e^{i\int A} q\;.
$$
In the supersymmetric theory both objects have the same string tension, so the Regge slopes are the same. In the orientifold theory, the gaugino maps to the antisymmetric fermion, so according to the equivalence the Regge slope of \eqref{baryon} and mesons should be the same.

We will study a \Ntwo superconformal version of the large-$N_c$ orientifold theory. The field theory can be constructed starting from $SU(N_c)$ \Nfour theory, adding flavor in the form of $N_f$ hypermultiplets in the fundamental representation and performing a $\bZ_2$ projection that preserves \Ntwo supersymmetry. For thorough discussions of \Ntwo theories similar to the one we consider, see \cite{Aharony:1998xz,Hong:2003jm,Kruczenski:2003be}. Before the projection, the field content can be arranged in \None fields as a vector multiplet, $W_\alpha$, three chiral multiplets $\Phi_1$,$\Phi_2$,$\Phi_3$ in the adjoint representation as well as $2 N_f$ chiral multiplets $Q^a$, $\Qt_a$ in the fundamental and anti-fundamental representation of the gauge group respectively. The covariant index $a=1,\dots,N_f$ refers to the fundamental representation of flavor. The superpotential has the form
$$
W=\sqrt{2}\tr \left([\Phi_1,\Phi_2]\Phi_3 \right)+Q^a\Phi_3\Qt_a+m_q\, Q^a\Qt_a
$$
There is an $SU(2)_L$ symmetry rotating $\Phi_1$ and $\Phi_2$ and a $SU(2)_R\times U(1)_R$ R-symmetry. The total global symmetry group is then
$$
SU(2)_R\times SU(2)_L\times U(1)_R\times U(N_f)
$$
Denoting $(j_R,j_L)_R$ as the spins $j_R$, $j_L$ of the representation under the $SU(2)_R$, $SU(2)_L$ groups and $R$ the $U(1)_R$ R-charge, supercharges are in the $(1/2,0)_1$ representation. The chiral fields $\Phi_1$, $\Phi_2$ are in the $(1/2,1/2)_0$ representation, the field $\Phi_3$ is in the $(0,0)_{\pm 2}$ representation and the flavor multiplets are in the $(1/2,0)_0$ representation.

In order to preserve supersymmetry, the $\bZ_2$ projection is done on the $SU(2)_L$ group. The effect of the projection is to change $\Phi_1$ and $\Phi_2$ into antisymmetric fields in color, $A^{[ij]}$ and ${\widetilde A}_{[ij]}$. The geometric interpretation in terms of a D-brane construction in string theory will be explained in section \ref{sec:model}. The chiral meson operators
\begin{equation}\label{bpsmeson}
\cM^a_b = Q^a \Qt_b , \ \ {\hat \cM}^a_b = Q^a \Phi_3 \Qt_b,
\end{equation}
are unaffected by this procedure, while it is not longer possible to build operators of the form $Q^a \Phi_{1,2} \Qt_b$. Instead, there are chiral operators similar to \eqref{baryon}
\begin{equation}\label{bpsbaryon}
\cB^{ab}= Q^a A Q^b\,.
\end{equation}
Since the chiral fields are bosonic operators, there are only two possibilities, either the operator is symmetric in flavor and the $Q$ operators are arranged in an antisymmetric representation of  $SU(2)_R$ or the reverse symmetric option. The lowest scalar component of the chiral primary operator that belongs to a short multiplet should satisfy the BPS condition $\Delta = 2 j_R+R/2$, where $\Delta$ is the conformal dimension of the operator. In both cases $\Delta=3$ and $R=0$, but for the symmetric flavor operator $j_R=1/2$ so it is not a BPS operator
as also explained in \cite{Aharony:1998xz}.
For the antisymmetric flavor operator there are two options, either $j_R=1/2$ or $j_R=3/2$, only the last one corresponds to a BPS operator.
 Consider now the operators most directly related to the baryons \eqref{baryon}
\begin{equation}\label{nonbpsbaryon}
B^{ab}=\psi_A \psi_Q^a \psi_Q^b\,
\end{equation}
where $\psi_X$ refers to the fermionic component of the chiral multiplet $X$. None of the fermionic fields are charged under $SU(2)_R$, so it should be in an antisymmetric representation of flavor. However, it does not belong to the BPS multiplet given by \eqref{bpsbaryon}, since the R-charge for fermionic components can be at most $R=\pm 1$ and this operator has $R=-3$. We will give the full spectrum of BPS scalar mesonic operators later, during the analysis of the holographic dual theory in section \ref{sec:mesons}.

Notice that the $\cM^a_b$ operators are in the adjoint representation of the $U(N_f)$ flavor group and are neutral under the $U(1)$ subgroup. On the other hand, the $\cB^{ab}$ operators are charged under the $U(1)$, which justifies the identification of mesons and baryons that we have assumed.
The only alternative baryons that are BPS are anti-symmetric in their flavor
quantum numbers and involve at least some scalar fields.
In particular, for $N_f=1$ there are no BPS baryons at all.
There
are also alternative baryons with fermions only or symmetric in flavor, but
they are not BPS, that is they are in long supermultiplets and
their mass is not protected against large corrections at strong coupling
as we will discuss in more detail in what follows.

\section{Holographic construction}\label{sec:model}

The brane construction that produces the supersymmetric orientifold theory is a special case of the setups studied in \cite{Park:1998zh}, based on the description of \Ntwo theories from D4 branes suspended between NS5 branes introduced in \cite{Witten:1997sc}. Here, we review their procedure for our case and compute explicitly the flavor group. Holographic models of non-supersymmetric orientifold theories have been considered using analogous brane constructions in type 0 theory, specifically as examples of theories that are conformal in the large $N_c$ limit, but also as equivalent to supersymmetric theories \cite{Klebanov:1999ch,Armoni:1999gc,Blumenhagen:1999uy,Angelantonj:1999qg,Dudas:2000sn,DiVecchia:2004ev}.

The construction in type IIA theory consists on a set of $2 N_c$ D4 branes wrapping a circle in the $x^6$ direction and intersecting two O6$^-$ planes at opposite sides of the circle. In addition, there is a NS5 brane at each orientifold point and $2 N_f$ D6 branes parallel to the O6 planes.
$$
\begin{array}{r|cccccccccc}
 \! & 0 & 1 & 2 & 3 & 4 &5 & 6& 7& 8 & 9 \\
\hline {\rm D4} & X & X & X & X& \cdot & \cdot & X & \cdot & \cdot & \cdot \\
{\rm O6/D6} & X & X & X & X & \cdot & \cdot & \cdot & X & X & X \\
{\rm NS5} & X & X & X & X & X & X & \cdot & \cdot & \cdot &\cdot
\end{array}
$$
The $N_f=4$ theory is conformal, the beta-function vanishes
identically. In the brane setup this corresponds to
the fact that all RR tadpoles cancel.  For $N_f \neq 4$ one has non-vanishing tadpoles
which result in a non-zero beta function for the 't Hooft coupling
$\lambda$ which is suppressed by $N_f/N_c$ at large $N_c$. So
to leading order in $N_f/N_c$ we can neglect the tadpoles and
consider the
D6s and O6s as probes just as in the D3/D7 system of \cite{Karch:2002sh}.
 The brane setup described so far also has T-dual
as a configuration of D3 and D7 branes. The two O6 planes map to a single O7 plane and the NS5 brane to a $\bZ_2$ singularity localized at $x^6=x^7=x^8=x^9=0$.
$$
\begin{array}{r|cccccccccc}
 \! & 0 & 1 & 2 & 3 & 4 &5 & 6& 7& 8 & 9 \\
\hline {\rm D3} & X & X & X & X& \cdot & \cdot & \cdot & \cdot & \cdot & \cdot \\
{\rm O7/D7} & X & X & X & X & \cdot & \cdot & X & X & X & X \\
\bZ_2 & X & X & X & X & X & X & \cdot & \cdot & \cdot & \cdot
\end{array}
$$
The geometric effect of the $\bZ_2$ action is a reflection in the transverse directions. The orientifold projection $\Omega'=\Omega R_{45} (-1)^{F_L}$ involves worldsheet parity reversal $\Omega$, a reflection $R_{45}$ in the $x^4$ and $x^5$ coordinates and $(-1)^{F_L}$ acts as -1 in the Ramond sector of left movers. The effect on Chan-Paton factors of open strings on D3 branes is given by the matrices
\begin{equation}
\gamma_3 = \left(\begin{array}{cc} i I_{N_c} & \\ &  -i I_{N_c} \end{array}\right)\,, \ \ \omega_3 = \left(\begin{array}{cc} & I_{N_c}  \\   - I_{N_c} & \end{array} \right)\,,
\end{equation}
where $I_{N_c}$ is the $N_c\times N_c$ identity matrix. The corresponding matrices for the D7 branes are
\begin{equation}
\gamma_7 = \left(\begin{array}{cc} i I_{N_f} & \\ &  -i I_{N_f} \end{array}\right)\,, \ \ \omega_7 = \left(\begin{array}{cc} & I_{N_f}  \\   I_{N_f} & \end{array} \right)\,,
\end{equation}

The massless spectrum of D3 branes involves a vector multiplet on the worldvolume $A_{0123}$ and three complex scalar multiplets describing the transverse motion $X_{45}$, $X_{67}$, $X_{89}$. The combined $\bZ_2$ and orientifold projection can be done as
\begin{equation}
\begin{array}{rclcl}
A_{0123} & \to & {\hat A}_{0123}=\gamma_3 \; A_{0123}\; \gamma_3^{-1} & \to & \hat{\hat A}_{0123} = - \omega_3\; {\hat A}_{0123}^T\; \omega_3^{-1} \\
X_{45} & \to & {\hat X}_{45}=\gamma_3 \; X_{45}\; \gamma_3^{-1} & \to&  \hat{\hat X}_{45} = - \omega_3\; {\hat X}_{45}^T\; \omega_3^{-1} \\
X_{67,89} & \to & {\hat X}_{67,89}=-\gamma_3 \; X_{67,89}\; \gamma_3^{-1} & \to&  \hat{\hat X}_{67,89} =  \omega_3\; {\hat X}_{67,89}^T\; \omega_3^{-1}
\end{array}
\end{equation}
The transformations of $A_{0123}$ and $X_{45}$ are identical and produce fields in the adjoint representation of $U(N_c)$. The projection on $X_{67,89}$ produces fields in a two-index supersymmetric antisymmetric representation. The resulting theory is a \Ntwo $U(N_c)$ theory with two antisymmetric hypermultiplets.

The D3/D7 spectrum is initially described by two $N_c\times N_f$ chiral multiplets $H^A$ describing strings from D3 to D7 branes and the reversed strings ${\widetilde H}_A=\epsilon_{AB} {H^B}^\dagger$. The projection acts as follows
\begin{equation}
\begin{array}{rclcl}
H^A & \to & {\hat H}^A=\gamma_3 H^A \gamma_7^{-1} &\to & {\hat{\hat H}_A}^* = i \epsilon_{AB} \omega_3 {\hat H}^B \omega_7^{-1}
\end{array}
\end{equation}
The massless field is a \Ntwo hypermultiplet in the $(N_c,\overline{N_f})$ representation.

The massless spectrum of D7 branes is split between vector fields in the 0123 and 6789 directions, $A_{0123}$ and $A_{6789}$, and a scalar field in the 45 directions, $X_{45}$. Transformations act in principle as
\begin{equation}\label{d7proj}
\begin{array}{rclcl}
A_{0123} & \to & {\hat A}_{0123}=\gamma_7 \; A_{0123}\; \gamma_7^{-1} & \to & \hat{\hat A}_{0123} = - \omega_7\; {\hat A}_{0123}^T\; \omega_7^{-1} \\
X_{45} & \to & {\hat X}_{45}=\gamma_7 \; X_{45}\; \gamma_7^{-1} & \to&  \hat{\hat X}_{45} = - \omega_7\; {\hat X}_{45}^T\; \omega_7^{-1} \\
A_{6789} & \to & {\hat A}_{6789}=-\gamma_7 \; A_{6789}\; \gamma_7^{-1} & \to&  \hat{\hat A}_{6789} =  -\omega_7\; {\hat A}_{6789}^T\; \omega_7^{-1}
\end{array}
\end{equation}
Since the 8d Poincar\'e invariance is broken in the worldvolume of the D7 branes the projection will be different for modes with dependence on the 6789 directions. The action \eqref{d7proj} for $A_{0123}$ and $X_{45}$ is valid for parity even modes while the action for $A_{6789}$ is valid for parity odd modes. This agrees with the $A_{0123}$ and $X_{45}$ components being scalar in the 6789 directions and $A_{6789}$ being a vector component. Parity odd modes are thus reduced to an element of $SO(N_f)$. From the point of view of the theory living on the D3 branes, these fields will correspond to chiral fields in an antisymmetric representation of the flavor group, so they should correspond to the baryon sector \eqref{bpsbaryon}. We will show this more explicitly in the discussion of the spectrum of modes on the D7 brane in the holographic dual.

The holographic dual description is type IIB string theory on $AdS_5\times \bR\bP^5$, with D7 probe branes that sit on top of O7 planes wrapping a $\bR\bP^3\subset \bR\bP^5$ cycle. The $AdS_5\times \bR\bP^5$ geometry can be understood using a different basis of transformations. The O7 action is $\Omega_7 = \Omega R_{45} (-1)^{F_L}$, while the $\bZ_2$ singularity acts as a $R_{6789}$ reflection on the geometry. Since O7 planes and O3 planes have the same effect on Ramond forms (cf.~\cite{Hanany:2000fq}), the combined action is equivalent to the action of an O3 plane $\Omega_3=R_{6789}\Omega_7 = \Omega R_{456789} (-1)^{F_L}$. The action of the O3 plane on $AdS_5\times S^5$ is known to give the $\bR\bP^5$ geometry, since it acts as a reflection on the space transverse to the D3 branes \cite{Witten:1998xy}. From the T-dual perspective this geometry without the O7 orientifold can be constructed from a stack of D4 branes sitting on O4$^-$ or O4$^+$, giving holographic duals with orthogonal or symplectic gauge groups.

In addition to the configuration we have considered here there are other
cases that give raise to the same geometry.
If the O6$^-$ are replaced by O6$^+$, the field in the
antisymmetric representation of color becomes a field in the
symmetric representation. One could also move the NS5 branes
away from the orientifolds in the $x^6$ direction,
giving $Sp(N_c)\times Sp(N_c)$ or $SO(N_c)\times SO(N_c)$
gauge theories with matter in the bifundamental representation
for O6$^-$ and O6$^+$ planes respectively. One can distinguish
different constructions studying the modes on the brane,
we will comment on this in the next section. For the readers convenience
we have summarized the various orientifold projections in
figure (1). Here cases a) and b) correspond to the theories studied in \cite{Witten:1997sc}, while
case c) is the model of \cite{Aharony:1998xz}. Cases a), b) and i) have a vanishing beta function for
$N_f=0$, c), e) and g) for $N_f=4$. All configurations give rise to a conformal theory in the probe limit.

\begin{figure}
\label{orientitable}
\hfill\includegraphics[scale=.53]{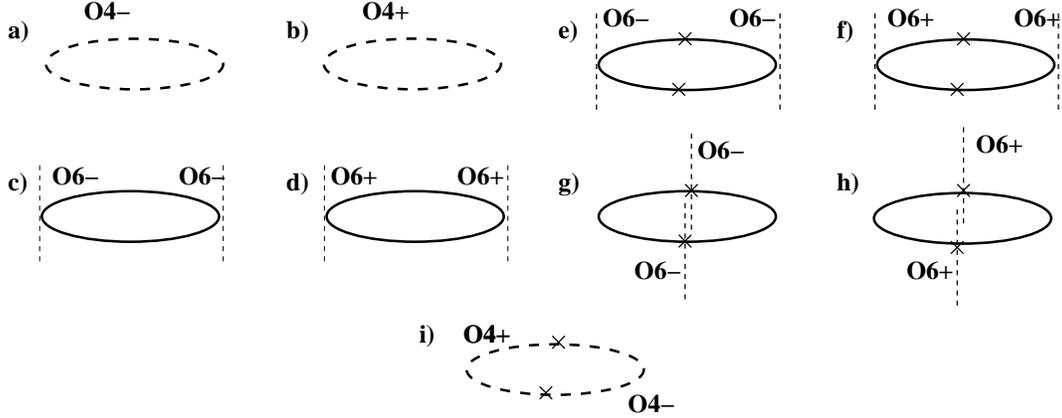}
\vspace*{-10pt}
\caption
  {Type IIA brane configurations dual to IIB setups with D3s, D7s and O7s
on a transverse $\bZ_2$ singularity. D4 branes wrap around the circle, dashed lines represent orientifold
planes and a cross represents NS5 branes.
a) and b) give rise to a theory
on $AdS_5 \times \bR \bP^5$
without O7s. The dual field theory has ${\cal N}=4$ supersymmetry
with $SO$ or $Sp$ gauge group respectively. c) and d) are dual to O7s
in $AdS_5 \times S^5$. The field theory has ${\cal N}=2$ supersymmetry
and Sp/SO gauge group with an antisymmetric/symmetric tensor hypermultiplet.
e) - h) represent O7s in  $AdS_5 \times \bR \bP^5$. e) and f) describe
$Sp \times Sp$ and $SO \times SO$ gauge groups with bi-fundamental matter,
g) and h) a single $SU$ gauge group with two anti-symmetric or symmetric
tensor hypermultiplets. g) is the theory described in detail in this
work as it realizes the alternative baryon scenario. The setup
i) gives an $SO \times Sp$ product gauge group. Additional
flavor branes can be added in all of the configurations.
}
\vspace*{-10pt}
\end{figure}

\section{Flavored spectrum}\label{sec:mesons}

Flavored states in the field theory are described as open string fluctuations on the D7 branes. There is a hierarchy of modes with masses proportional to different powers of the string coupling and the string length that translate into $N_c$ and the 't Hooft coupling. In terms of the quark mass, the lightest states $\sim m_q/\sqrt{\lambda}$ correspond to fluctuations of massless modes on the D7 brane, that can be mapped to BPS operators like (\ref{bpsmeson},\ref{bpsbaryon}).
The next level are small open strings attached to the D7 brane $\sim m_q/\lambda^{1/4}$, that can be identified as non-BPS operators like \eqref{nonbpsbaryon}. Highly excited states or very large operators create states with energies $\sim m_q$ and map to large strings that can be described using a semiclassical approximation with the classical Nambu-Goto action.

Heavy flavored states, that is states whose mass grows proportional to $N_c$, are described by wrapped branes with strings joining them to the flavor branes. Heavy baryons are D5 branes wrapping the $\bR\bP^5$ with $N_c$ strings attached to it \cite{Witten:1998xy}. The mass of the baryon scales as $\sim N_c m_q$. It is worth noting that the five-from flux on the $\bR\bP^5$ is actually $N_c/2$, but a consistent D5 brane configuration has to wrap twice. This implies that, contrary to geometries with a $S^5$ factor in the metric, the D5 brane is not a topologically stable object, although it can be dynamically stable. From the field theory perspective this can be easily understood. In a $SU(N_c)$ theory where all the two-index fields are in the adjoint representation, the baryon carries a conserved $U(1)_B$ charge, so it cannot decay to lower states like mesons, that are neutral. However, in the orientifold theory there are lighter states that are also charged under $U(1)_B$, those are the light baryons like \eqref{bpsbaryon} that we have been discussing.

A consistent identification of Pfaffian mesons \eqref{meson} are D3 branes wrapping the $\bR\bP^3$ cycle, with masses $\sim N_c m_q/\sqrt{\lambda}$. The same kind of arguments as for the Pfaffian of orthogonal gauge group explained in \cite{Witten:1998xy} apply to this case. The existence of a Pfaffian is related to the presence of discrete fluxes on $\bR\bP^5$. There can be fractional 2-form flux for Ramond $\theta_R$ or Neveu-Schwarz fields $\theta_{NS}$ due to the non-trivial twisted homology $H_2(\bR\bP^5,\widetilde \bZ)=\bZ_2$. Our construction relies on the orientifolds being of the O6$^-$ type, so there is no NS flux $\theta_{NS}=0$. This is consistent, since a $\theta_{NS}\neq 0$ will forbid the wrapping of the D3 brane. As summarized in figure (1), replacing in our case the O6$^-$ with O6$^+$ we replace the anti-symmetric hypermultiplets with symmetric hypermultiplets. Indeed this theory should not have a Pfaffian meson; the wrapped D3 is forbidden by the NS flux which is non-zero for the O6$^+$ type brane. The R flux is related to the rank of the group, $\theta_R=0$ corresponds to $N_c$ even and $\theta_R\neq 0 $ to $N_c$ odd. This also agrees with our interpretation. In the $\theta_R\neq 0$ case there is an induced charge on the D3 brane from the Chern-Simons coupling to the Ramond $C_2$ form
\begin{equation}
\int C_2 \wedge F_2
\end{equation}
The D3 brane can be wrapped if a string is attached to the D3 brane to cancel the total charge. In the field theory this corresponds to contracting a fundamental field with the Pfaffian operator, that is the situation for $N_c$ odd gauge group in \eqref{meson}.

We will show now explicitly that light baryons map to small fluctuations of the flavor branes, in the same way mesons do, by identifying the modes that are associated to scalar BPS operators. The spectrum of massless scalar excitations of the unprojected theory was studied in \cite{Kruczenski:2003be}. There is a Kaluza-Klein tower of modes labeled by the angular momentum $\ell$ on the $S^3$ the D7 brane wraps. The isometry group of the $S^3$ is $SO(4)\simeq SU(2)_R\times SU(2)_L$. There is an additional $U(1)_R$ group associated to rotation on the plane transverse to the D7 branes. Supersymmetries of the background are in a $(j_R,j_L)_R=(1/2,0)_1$ representation and modes fall into four dimensional hypermultiplets whose lowest component is in a $\left({\ell \over 2}+1, {\ell \over 2}\right)_0$ representation. Using the holographic dictionary, the conformal dimension associated to each component can be read from the mass of the modes. Then, the bosonic components of the multiplet can be split as follows:
\begin{itemize}

\item[i)] The lowest $\Delta =\ell+2$ component is a spin -1 mode of the D7 vector field in the $S^3$ directions with $\ell+1$ angular momentum, $A_-^{\ell+1}$.

\item[ii)] The $\Delta =\ell+3$ component divides into $\left({\ell\over 2},{\ell\over 2} \right)_{\pm 2}$ and $\left({\ell\over 2},{\ell\over 2} \right)_0$ contributions. The $R=\pm 2$ contribution corresponds to a mode of the transverse scalar field with $\ell$ units of angular momentum, $\Phi^\ell$. The $R=0$ component belongs to the vector field in the D7 brane directions transverse to the $S^3$, also with $\ell$ angular momentum, $A^\ell$.

\item[iii)] The $\Delta=\ell+4$ component is the spin +1 mode with $\ell-1$ units of angular momentum of the vector field in the $S^3$ directions, $A_+^{\ell+1}$.

\end{itemize}
The fermionic components of the multiplet are four-dimensional Dirac spinors with a chirality associated to the $S^3$ that derives from the decomposition of ten-dimensional spinors in the full geometry \cite{Kirsch:2006he}. We can distinguish two different kind of components
\begin{itemize}
\item[i)] A fermion with angular momentum $\ell$ and `right' chirality under the $SO(4)$ group $\Psi^+_\ell$,
\item[ii)] A fermion with angular momentum $\ell-1$ and `left' chirality under the $SO(4)$ group $\Psi^-_{\ell-1}$
\end{itemize}

Let us see now how the projection affects to the fields in the multiplet. If the $S^3$ is embedded in $\bR^4$ as
$$
x_1^2+x_2^2+x_2^2+x_4^2=1\,,
$$
the $\bZ_2$ projection that changes the $S^3$ into an $\bR\bP^3$ space can be seen as the reversal $x_i\to -x_i$, $i=1,2,3,4$. Then, the action over the fields will depend on their angular momentum and their spin
\begin{equation}
\begin{array}{lclclc}
\Phi^\ell \to (-1)^\ell \Phi^\ell & , & A^\ell \to (-1)^\ell A^\ell &, & A_\pm^\ell \to (-1)^{\ell+1} A_\pm^\ell \,.
\end{array}
\end{equation}
In terms of $SU(2)_L\times SU(2)_R$, the reflection can be seen as the action of a $\bZ_2 \subset SU(2)$ center element over the $2\times 2$ matrix $\bx = x_i\sigma^i$, where $\sigma^i$ are the Pauli matrices and the identity and a general transformation acts as $\bx \to U_L \bx U_R^\dagger$. There are two possible actions over fermions, depending on whether the center group belongs to the `left' or `right' group. For $\bZ_2\subset SU(2)_L$,
\begin{equation}
\begin{array}{lclc}
\Psi^+_\ell \to (-1)^\ell\Psi^+_\ell &, & \Psi^-_\ell \to (-1)^{\ell+1}\Psi^-_\ell
\end{array}
\end{equation}
Notice that this action preserves supersymmetry, since all the components in the same multiplet transform in the same way. We have included fermionic fields in the discussion to show this explicitly, but we will concentrate on the bosonic components in the following.

In principle multiplets with odd $\ell$ would be projected out from the spectrum, but the geometric action on the D7 brane can be complemented with an action over the matrix indices of the field. Since the matrices are in the adjoint representation of $SU(N_f)$, transposition is the only independent transformation that squares to unity. For $\ell=2n+1\geq 1$,
\begin{equation}\label{projd7}
\begin{array}{lclcl}
\Phi^{2n+1} \to - (\Phi^{2 n+1})^T & , & A^{2 n+1} \to - (A^{2 n+1})^T &, & A_\pm^{2 n+2} \to -( A_\pm^{2 n+2})^T\,.
\end{array}
\end{equation}
With this choice, $\ell$ even modes are untouched and $\ell$ odd modes are in an antisymmetric representation of flavor. This is clearly the projection corresponding to the theory we are interested in, that is case g) from figure 1. The  $\ell =0$ mode can be matched to the BPS mesons \eqref{bpsmeson}
\begin{equation}
\cM^a_b = Q^a \Qt_b , \ \ {\hat \cM}^a_b = Q^a\, X_{67,89}\, \Qt_b,
\end{equation}
where the scalar component of $\cM$ corresponds to $A_-^1$ and the one of ${\hat \cM}$ to $\Phi^0$. Similarly, the lowest $\ell=1$ mode $A_-^2$ matches with the scalar component of the BPS baryon \eqref{bpsbaryon}
\begin{equation}
\cB^{[ab]}= Q^a X_{45} Q^b\,.
\end{equation}

The action \eqref{projd7} is not unique, modes with even angular momentum could also be projected, so for all $\ell$
\begin{equation}\label{ortproj}
\begin{array}{lclcl}
\Phi^\ell \to (-1)^\ell (\Phi^\ell)^T & , & A^\ell \to (-1)^\ell (A^\ell)^T &, & A_\pm^\ell \to (-1)^{\ell+1}( A_\pm^\ell)^T\,.
\end{array}
\end{equation}
In this case even $\ell$ modes map to a symmetric representation of flavor while odd $\ell$ modes are antisymmetric. It is easy to see that this gives the right spectrum for a theory with $SO(N_c)$ gauge group, with operators of the form
\begin{equation}
\cM^{(ab)}= \delta^{ij} Q_i^a Q_ j^b, \ \ \ \cM^{[ab]}= X_{45}^{[ij]} Q_i^a Q_j^b
\end{equation}
where $i,j$ are color indices and $X$ is a field in the adjoint of $SO(N_c)$. This corresponds to the O4$^-$ construction we mentioned in section \ref{sec:model}; case a) from figure (1).

Theories with symplectic group can also be found in a similar way. We have commented before that it could be possible to have discrete NS flux on $\bR\bP^5$. As was explained in \cite{Witten:1998xy}, the path integral of strings is modified by a factor
$$
\exp\left(i \int_{\bR\bP^2} B_2\right) =-1
$$
so the transformations \eqref{ortproj} should pick up this sign. This means that symmetric and antisymmetric representations are interchanged, and the spectrum matches with the one of a $Sp\,( N_c)$ theory
\begin{equation}
\cM^{[ab]}= J^{[ij]} Q_i^a Q_ j^b, \ \ \  \cM^{(ab)}=X^{(ij)} Q_i^a Q_j^b
\end{equation}
with
$$
J=\left(\begin{array}{cc} & I_{N_c/2} \\ -I_{N_c/2} & \end{array} \right)\,.
$$

The case with symplectic gauge group corresponds to the construction with O4$^+$ planes (case b) from figure 1), so the introduction of fractional NS flux can be seen as the change of O4$^-$ to O4$^+$. The same can be applied to the O6 cases, but in this case the breaking of 8d Poincar\'e invariance on the D7 brane worldvolume means that only the projection over odd $\ell$ modes change sign in \eqref{projd7}
\begin{equation}\label{projd7b}
\begin{array}{lclcl}
\Phi^{2n+1} \to  (\Phi^{2 n+1})^T & , & A^{2 n+1} \to  (A^{2 n+1})^T &, & A_\pm^{2 n+2} \to ( A_\pm^{2 n+2})^T\,.
\end{array}
\end{equation}
and the $\ell$ even modes do not change. This gives the right spectrum for a theory with a symmetric hypermultiplet,
case h) from figure (1),
\begin{equation}
\cM^a_b = Q^a \Qt_b , \ \ \ \cB^{(ab)}= X_{45}^{(ij)} Q_i^a Q_j^b
\end{equation}

When the NS5 branes are not stuck at the O6 planes (that is cases e) and f) from figure (1)), the geometric action is different. Now the modes that are projected out are the even $\ell$ modes. For O6$^-$ ($\theta_{NS}=0$) the even modes are in the antisymmetric representation of flavor, while for O6$^+$ ($\theta_{NS}\neq 0$) the modes are in the symmetric representation. Odd $\ell$ modes stay in the adjoint representation. This agrees with the expectation for a $Sp\,(N_c)\times Sp\,(N_c)$ theory in the O6$^-$ case and a $SO(N_c)\times SO(N_c)$ theory in the O6$^+$ case.
\begin{equation}
\cM_{\rm Sp}^{[ab]}= J^{[ij]} Q_i^a Q_ j^b, \ \ \ \cM_{\rm SO}^{(ab)}= \delta^{ij} Q_i^a Q_ j^b\,.
\end{equation}
This exhausts all possible configurations based on two NS5 branes and two O6 planes of the same kind. We have seen that there is a nice correlation with the possible choice of discrete fluxes in the $AdS_5\times \bR\bP^5$ geometry.

\section{Conclusions}

We have given a holographic example of large-$N_c$ equivalences between supersymmetric gauge theories. The equivalence relates theories that can have $SO(N_c)$ or $Sp\,(N_c)$ adjoint fields and $SU(N_c)$ theories with matter in the antisymmetric or symmetric representation. These theories are all described by the same geometry, but different topological configurations on the geometry lead to a different spectrum of gauge invariant operators in the uncommon sectors.

Large N equivalence predicts that all states in the parent theory which are invariant under the $\bZ_2 \times  \bZ_2$ projection have a corresponding
state in the daughter theory with the same mass in the large $N_c$ limit. This is clearly true in our example, since the states that survive the
orientifold description have an identical description on the supergravity side in both parent and daughter. For instance, the BPS baryonic
operators $A_{[ij]} Q^i Q^j$ as well as the BPS mesonic operators $X_i^j Q^i \widetilde Q_j$ of the ${\cal N}=2$ theory have the same mass as the
operators $X_i^j Q^i \widetilde Q_j$ of the ${\cal N}=4$ theory since both correspond to the same kind of small fluctuations of the probe flavor
brane in the holographic dual.

The analysis of which subset of the BPS states of the ${\cal N}=4$ theory survives the orientifold projection is identical to the analysis of which of the dual supergravity modes survives the orientifold projection. After all, they have the same quantum numbers under all global symmetries.
Another object whose properties are inherited is the tension of a long flux tube. Again, this is obviously true as the description of flux tubes is identical in both theories.

Both light mesons and baryons correspond to open string fluctuation on the flavor branes, with differences of energy that are of order $\sim m_q/\sqrt{\lambda}$. For high excitations $\sim m_q$ described by semiclassical strings, it is not possible to distinguish small differences in quantum numbers, so light baryons and mesons have the same high energy spectrum. For the model we have discussed, there is a larger degeneracy between baryon and mesons with different radial excitation number $n$ due to an enhanced $SO(5)$ symmetry. As was observed in \cite{Kruczenski:2003be}, modes with the same $n+\ell$ have the same mass, so a baryon  state defined by $n$, $\ell$ is degenerate with other states $n'$, $\ell'$ that are baryons if the differences $n-n'$, $\ell-\ell'$ are even or mesons if the differences are odd.

We have also shown that in theories with alternative baryons
we have mesons and baryons coexisting in a rich hierarchy of large $N_c$
scalings for the masses.
We find both conventional baryons with masses of order $N_c$ and alternative
baryons with masses of order 1. Similarly, there are standard mesons
and Pfaffian mesons with masses of order 1 and $N_c$ respectively.
At strong coupling the low-spin altenative baryons split into BPS baryons with masses of order
the meson mass as well as non-BPS baryons with masses $\lambda^{1/4}$
above the meson mass.

Most of our analysis has been topological, the $AdS_5$ part of the geometry playing no role. We expect then that the same kind of arguments should apply to holographic duals of confining theories. Since long flux tubes in the field theory are described as large strings in the holographic dual, light mesons and baryons will have the same Regge slope.

\section*{Acknowledgments}
We would like to thank A. Armoni and A. Patella for useful discussions. We also would like to thank University of Washington's Institute for Nuclear Theory where this work was initiated during the ``New frontiers in large N Gauge Theories'' workshop. This work was supported in part by the U.S. Department of Energy under Grant No. DE-FG02-96ER40956.

\end{document}